\documentclass[fleqn,12pt,twoside]{article}
\usepackage[headings]{espcrc1}

\readRCS
$Id: espcrc1.tex,v 1.2 2004/02/24 11:22:11 spepping Exp $
\usepackage{graphicx}
\usepackage[figuresright]{rotating}

\newcommand{\be}{\begin{equation}}
\newcommand{\ee}{\end{equation}\noindent}
\newcommand{\bea}{\begin{eqnarray}}
\newcommand{\eea}{\end{eqnarray}}
\newcommand{\la}{\langle}
\newcommand{\ra}{\rangle}

\newcommand{\AmS}{{\protect\the\textfont2
  A\kern-.1667em\lower.5ex\hbox{M}\kern-.125emS}}

\title{
Viscosities of Hot Gluon -- A Lattice QCD Study --
}

\author{
	A. Nakamura\address[Hiroshima]{RIISE, Hiroshima university,
        Higashi-Hiroshima 739-8521, Japan}
	and
        S. Sakai\address[Yamagata]{Yamagata University, 
	Yamagata 990-8560, Japan}
}
       
\begin{document}

\maketitle

\begin{abstract}
We present transport coefficients (shear viscosity, $\eta$, and bulk 
viscosity, $\zeta$) for the gluon system obtained by the lattice QCD.
This is an indispensable calculation towards the understanding of
``New State of Matter'' observed in RHIC.
We study the temperature regions of RHIC ($1.4 \leq T/T_c \leq 1.8$)
and much higher ones up to $ T/T_c \sim 20$. 
In RHIC regions, the ratio of shear viscosity to entropy
density, $\eta/s$, is around  $\sim 0.1-0.4$, and satisfies the KSS 
bound.
At high temperature, $\eta$ becomes two or three oder of magnitude
larger.

Our calculation has two limitations: (i) the use of the quench approximation,
i.e., without quark pair creation-annihilation effects on vacuum,
and (ii) the use of an ansatz for the spectral function. 
The first point has been well studied in calculations of the spectroscopy and 
the phase-transition behavior.
To investigate the second point, we compare our results with
perturbative calculations in high $T$-regions, and also check
the effects of the modification of the spectral function
on the viscosity.
\end{abstract}

\section{INTRODUCTION -- Matter in deconfinement region}

More than twenty years ago,
Gross, Pisarski and Yaffe\cite{GrossPisarkiYaffe81}
wrote as follows:
``Now that we possess a theory of the strong interactions,
it is natural to explore the properties of hadronic matter in unusual
environments,  in particular at high
temperature or high baryon density.
There are three places where one might look
for the effects of high temperature and/or
large baryon density, (1) the interior of neutron stars,
(2) during the collision of heavy ions at very high energy per nucleon,
and (3) about $10^-5$ sec after the big bang''.
We are now in an excited era. At RHIC, the confinement/deconfinement
transition temperature is probably exceeded, and for the first time 
in science history, a deconfinement system is created in a laboratory
on earth. 

The outcome is very surprising:  The matter produced does not look as 
a quasi-free gas as naively expected, but rather is well described as 
a fluid.  In SPS energy regions, the hydro-model describes
well one-particle distributions, HBT etc., but fails to describe
the elliptic flow data.
This may be not so surprising. Fifty years ago, Landau criticized
Fermi's statistical model\cite{Fermi}, and noticed
`owing to high density of the particles and to strong interaction
between them, one cannot really speak of their number'  
and proposed his relativistic hydro-dynamical model\cite{Landau}.
The first quantum field theoretical analysis of the applicability
conditions of the Landau hydro-dynamical model was reported 
in Ref.\cite{Namiki}.

In three-dimensional hydro-dynamical calculations to analyze RHIC data,  
it is assumed that the matter produced is a perfect fluid, 
i.e., {\it its viscosity is zero.} 
This assumption is supported by several phenomenological analyses.
This also suggests that the new state of matter produced at RHIC should be
treated as a strongly coupled system:  The perturbative calculation
results in
\be
 \eta = \frac{\eta_1 \cdot T^{3}}{g^{4} \ln(\mu^{*}/g T)}.
\label{eta-perturbation}
\ee
The viscosity is small when $g$ is large.
This is understandable because there should be sufficient frequent momentum
exchange to realize a perfect fluid.
Policastro, Son and Starinets have shown an example of a
strongly coupled
theory in which the viscosity is indeed very small, 
i.e., $\eta/s=1/4\pi$\cite{PTS01,Starinets05}.
They stressed that this is much smaller than that of ordinary matter, such as 
water or liquid helium, and conjectured that this is the lowest
bount (KSS bound)\cite{KSS04}.

It is thus important now to calculate the transport coefficients from
QCD, non-perturbatively.

\section{Transport Coefficients on lattice}

On the lattice, the calculation of the transport
coefficients is formulated in the framework of the linear response theory
\cite{Zubarev,Horsley}.
\begin{eqnarray}
\eta 
= - \int d^{3}x' \int_{-\infty}^{t} dt_{1} e^{\epsilon(t_{1}-t)}
     \int_{-\infty}^{t_{1}}
     dt' \la T_{12}(\vec{x},t)T_{12}(\vec{x'},t')\ra_{ret} .
\label{off-diagonal}
\end{eqnarray}
Here, $\la T_{\mu \nu} T_{\rho \sigma}\ra_{ret}$ is the 
retarded Green's function
of energy momentum tensors $T_{\mu \nu}$ at a given temperature. 
In the quenched approximation,
the energy momentum tensors are constructed from only gluonic field strength terms.
Bulk viscosity is defined in a similar manner.

Shear viscosity in Eq.(\ref{off-diagonal}) is also expressed 
using the spectral function $\rho$ of the retarded Green's 
function $\rho(\omega)$\cite{Horsley} as
\begin{eqnarray}
\eta = \pi \displaystyle{\lim_{\omega \rightarrow 0}}
 \frac{\rho(\omega)}{\omega}
= \pi\lim_{\omega \rightarrow 0} \frac{d \rho(\omega)}{d \omega}.
\label{diff}
\end{eqnarray}
It is determined by the shape of the spectral function near $\omega=0$.

For evaluating $\rho(\omega)$,
we use a well known fact that the spectral function of the retarded
Green's function at temperature $T$ is the same as that of Matsubara-Green's
function. Therefore, our target is to calculate  Matsubara-Green's
function($G_{\beta}(t_n)$) on a lattice and determine $\rho$ from 
it\cite{HNS93}.

To determine the spectral function $\rho(\omega)$ from
$G_{\beta}(t_n)$, 
we adopt the simplest non-trivial ansatz, i.e.,
a Bright-Wigner type ansatz
proposed by Karsch and  Wyld\cite{Karsch},
\bea
\rho_{BW}(\omega)
 = \frac{A}{\pi}(\frac{\gamma}{(m-\omega)^2+\gamma^2}-
 \frac{\gamma}{(m+\omega)^2+\gamma^2})
\label{rho_ka}
\eea
As this formula has already 3 parameters, to determine them,
the lattice size in temperature direction($N_T$) should be
$N_T \geq 8$. Thus, the minimum lattice size should be
$24^3 \times 8$, to obtain non trivial results.

Simulations are carried out using the Iwasaki's improved action and
standard Wilson action.
The simulations are performed at $\beta=$3.05,
3.3, 4.5 and 5.5 for the improved action 
and at $\beta$=7.5 and 8.5 for Wilson
action.  With roughly $10^6$ MC measurements at each $\beta$, we
determine Matsubara-Green's functions $G_{\beta}(t_n)$.
The errors of $G_{\beta}$ are still large in the large $t$ region,
however, we fit them with the spectral function $\rho$ given by
Eq.(\ref{rho_ka}).

The bulk viscosity is equal to zero within the range of error bars,
whereas the shear viscosity remains finite.

\begin{figure}[bt]
\begin{center}
\begin{minipage}{ 0.45\linewidth}
\includegraphics[width=1.00\linewidth]{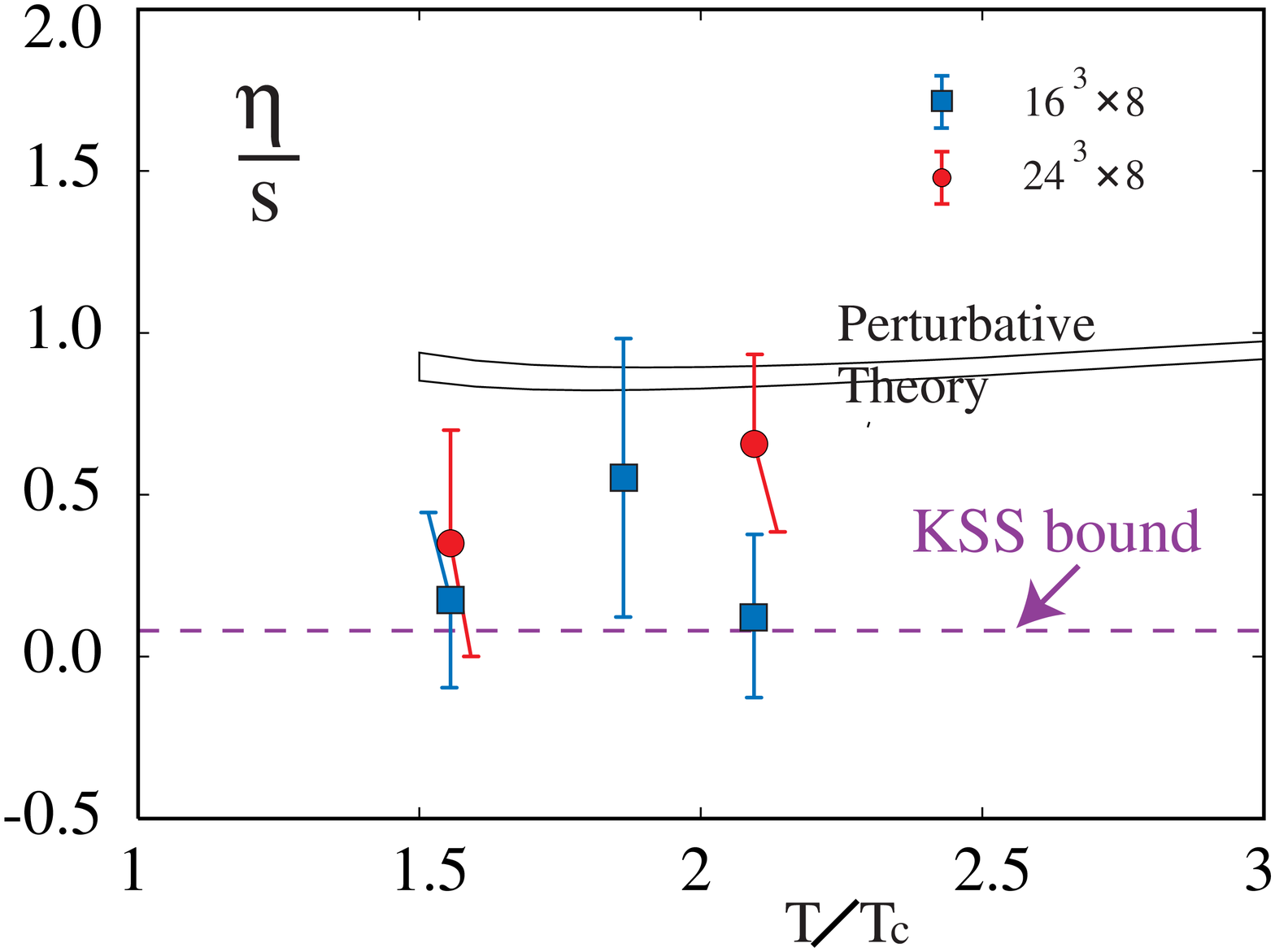}
\end{minipage}
\hspace{1mm}
\begin{minipage}{ 0.45\linewidth}
\includegraphics[width=1.00\linewidth]{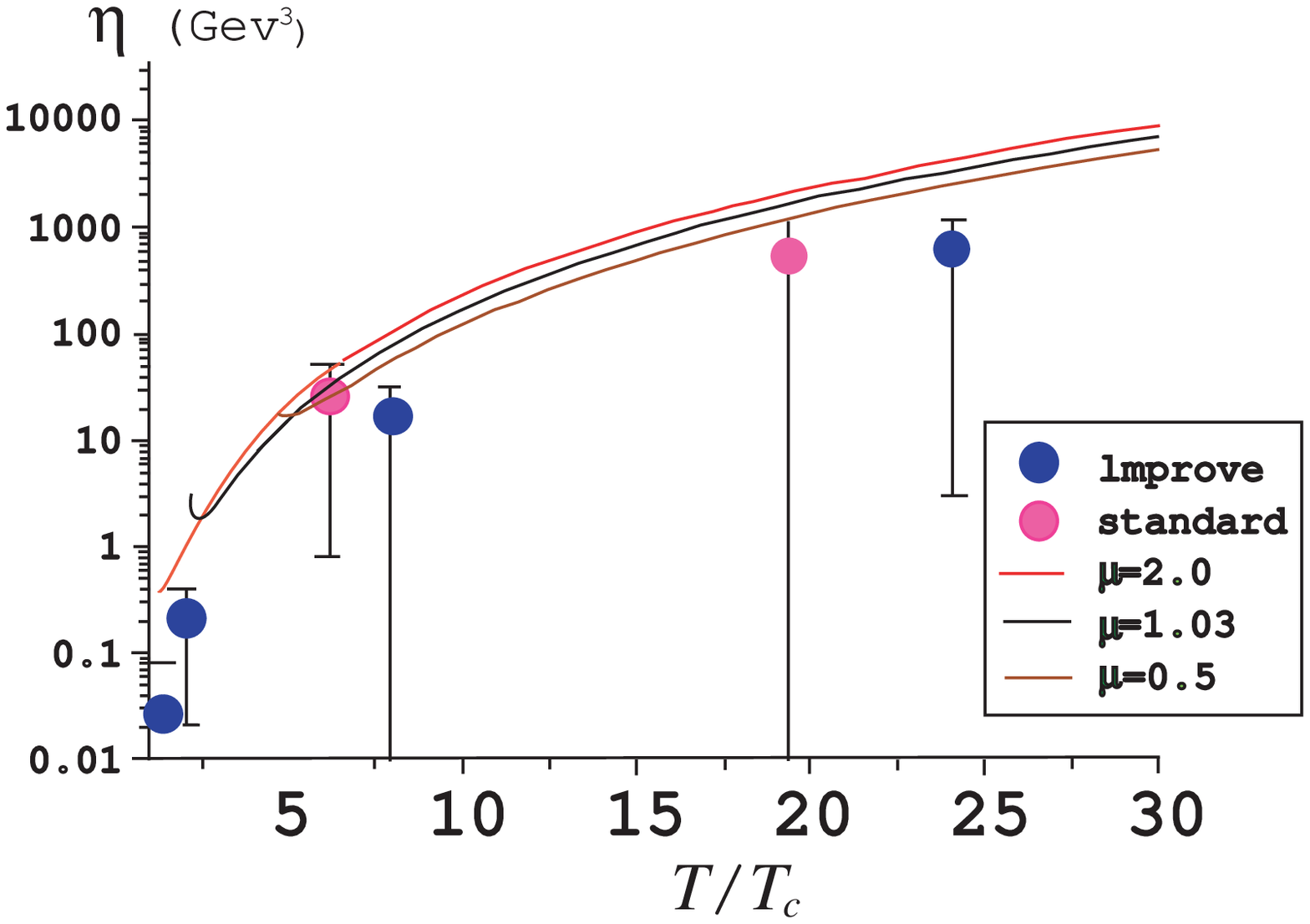}
\end{minipage}
\end{center}
\vspace{-3mm}
\caption{Shear viscosity as a function of temperature.  
Its ratio to entropy density in
RHIC temperature regions (left) and data in physical unit 
in wide temperature regions together with
perturbative calculations (right). 
}
\vspace{-3mm}
\label{eta}
\end{figure}

\section{Conclusions}

We may compare our results with the perturbation results
of $\eta$ in rather high temperature regions.
In perturbation,
bulk viscosity becomes zero\cite{Horsley,Kajantie},
whereas
shear viscosity in the next-to-leading log
is given by Eq.(\ref{eta-perturbation}). As seen in the right-hand 
figure of Fig.1,
 in low-$T$ regions, the perturbative calculation becomes inapplicable. 
At very high temperature, lattice and perturbative results
are satisfactorily consistent with each other. 
Although our result depends on the
assumption regarding $\rho_{BW}$ given in Eq. \ref{rho_ka}, it may be a
reasonable approximation of $d\rho/d\omega$ at $\omega=0$.

Aarts and Resco has proposed an another form of 
$\rho$ as \cite{Aarts}
\bea
 \rho(\omega)=\rho^{low}(\omega) + \rho^{high}(\omega),
\label{rho_a}
\eea
\begin{eqnarray}
\rho^{high}(\omega)=\theta(\omega-4m_{th}^2) \frac{d_A (\omega^2-4m_{th}^2)^{5/2}}
{80 \pi^2 \omega }[n(\omega/2)+0.5],
\label{rho_high}
\end{eqnarray} 
where $d_A=N_c^2-1$  and $n(\omega)=1/(\exp(\omega/T)-1)$.
$\rho^{low}(\omega)$ is a rational function with coefficients
as parameter.
 
In order to study the effect of $\rho^{high}$ on the 
shear viscosity, $\eta$, we assume that $\rho$ is given by 
$\rho=\rho_{BW}+\rho^{high}$, where $\rho_{BW}$ is given by Eq.(\ref{rho_ka}).
By changing $m_{th}$, the change in  $\eta$ is studied at $\beta=3.3$ of
improved action. When $\rho^{high}$ is absent( $m_{th}=\infty$), $\eta
a^3$=0.00225(201). If $m_{th}$ is set to be 5.0, 3.0 and 2.0, $\eta a^3$ becomes
0.00223(0.00191), 0.00194(0.00194) and 0.00126(0.00204), respectively.
At $m_{th}=1.8$, the contribution from $\rho^{high}$ becomes larger than
$G_{\beta}(t_n)$ of simulation at $t=1$, that fit could  not be done.
Generally, as $m_{th}$ decreases, the contribution from $\rho^{high}$ increases
and $\rho$ in the small $\omega$ region is suppressed. In this 
case, it results in a
decrease in $\eta$. 

We have calculated Matsubara-Green's function
and determine the shear viscosity of gluon plasma.  
In the high-temperature region, the agreement of the lattice 
and perturbative calculation is satisfactory. 
The lattice result of $\eta/s$ in
$T/T_c \le 3$ is smaller than that obtained by 
the extrapolation of the perturbative
calculation and satisfies the KSS bound.
From the well known relation between the 
mean free path and viscosity, our results also suggest that gluon
plasma is strongly interactive. 

Although our results depend on the
form of the spectral function
$\rho_{BW}$ given by Eq.(\ref{rho_ka}), we think that the qualitative 
features will
not change,  because as discussed, our
results are stable
if the high frequency part of the spectral function is included. 
We think that $\eta$  and $\eta/s$ will not reach 10 times of the
present value when more accurate determination of the transport
coefficients is carried out.

However, it is important to carry out a more reliable and accurate
calculation of transport coefficients, independent of the assumption
regarding the spectral function. 
To this purpose, we are starting the simulation
on an anisotropic lattice, to apply maximum entropy method.


\begin{thebibliography}{99}

\bibitem{GrossPisarkiYaffe81}:
        D.~J.~Gross, R.~D.~Pisarski and  L.~G.~Yaffe,
        \emph{Rev. Mod. Phys.}  {\bf 53}, 43 (1981).

\bibitem{Fermi} E. Fermi, \emph{Prog. Theor. Phys.} {\bf 5} (1950) 570.
 
\bibitem{Landau} S. Z. Belen'ski and L. D. Landau,
	\emph{Nuovo. Cimento Suppl.} {\bf 3} (1956) 15. 

\bibitem{Namiki} C. Iso, K. Mori and M. Namiki,
	\emph{Prog. Theor. Phys.} {\bf 22} (1959) 403.

\bibitem{PTS01} G. Policastro, D.T.Son and A.O.Starinets,
        Phys. Rev. Lett. 87 (2001) 081601,.
        (hep-th/0104066).

\bibitem{Starinets05} A.O.Starinets, in these Proceedings.

\bibitem{KSS04} P. Kovtun, D.T.Son and A.O.Starinets,
        hep-th/0405231.

\bibitem{Zubarev} D.N.~Zubarev,
        \emph{Non-equilibrium statistical mechanics}, Plenum, New York,
        1974

\bibitem{Horsley} R.~Horsley, W.~Schoenmaker,
        \emph{Quantum Field Theories out of Thermal Equilibrium,
        (I).General considerations, (II).
        The transport coefficients for QCD} \emph{Nucl. Phys.} {\bf
        B280[FS18]},716, 735(1987)

\bibitem{Kajantie} A.~Hosoya and K.~Kajantie, \emph{Transport
        Coefficients of QCD Matter} \emph{Nucl. Phys} {\bf B250}, 666(1985)

\bibitem{HNS93} T.Hashimoto, A.Nakamura and I.O.Stamatescu,
             \emph{Nucl. Phys.} {\bf B400}, (1993) 267.

\bibitem{Karsch} F.~Karsch and H.W.~Wyld,\emph{Thermal Green's function
	and transport coefficients on the lattice} \emph{Phys. Rev.}
	{\bf D35}, 2518(1987)



\bibitem{Arnold} P.~Arnold, G.D.~Moore and G.~Yaffe,
        \emph{Transport coefficients in high temperature gauge theories:
        (II) Beyond leading log}
        \emph{JHEP}{\bf 05} 051(2003),(hep-ph/0302165)



\bibitem{Aarts} G.~Aarts and J.M.M.~Resco, \emph {Transport
	coefficients: spectral function and the lattice} \emph{JHEP} {\bf
	4}, 53(2002)

\end{thebibliography}
\end{document}